\documentclass[]{spie}  

\usepackage{amsmath,amsfonts,amssymb}
\usepackage{siunitx}
\DeclareSIUnit{\gee}{\textit{g}}
\DeclareSIUnit{\bps}{bps}
\usepackage{graphicx}
\usepackage{subcaption}
\usepackage[colorlinks=true, allcolors=blue]{hyperref}

\title{The filter exchange system of the LSSTCam at \newline the Vera C. Rubin Observatory}

\author[a]{Alexandre~Boucaud}
\author[b]{Pierre~Antilogus}
\author[c]{\'{E}ric~Aubourg}
\author[d]{Antoine~Bernard}
\author[d]{Johan~Bregeon}
\author[e]{Patrick~Breugnon}
\author[b]{Julien~Cordian}
\author[f]{Hervé~Croizet}
\author[b]{Guillaume~Daubard}
\author[g]{Kevin~Fanning}
\author[e]{Fabrice~Gallo}
\author[g]{Anthony~S.~Johnson}
\author[b]{Claire~Juramy-Gilles}
\author[e]{Pierre~Karst}
\author[d]{Mile~Kusulja}
\author[d]{Eric~Lagorio}
\author[g]{Travis~Lange}
\author[b]{Didier~Laporte}
\author[g]{Juan-Carlos~Lazarte}
\author[g]{Margaux~Lopez}
\author[e]{Aur\'{e}lien~Marini}
\author[g]{Stuart~Marshall}
\author[g]{Dmitry~Onoprienko}
\author[g]{Hannah~M.~M.~Pollek}
\author[f]{Max~Turri}
\author[h]{Yousuke~Utsumi}
\author[d]{Francis~Vezzu}
\author[a]{Françoise Virieux}
\author[e]{T\'{e}o~Weicherding}

\affil[a]{Universit\'e Paris Cit\'e, CNRS/IN2P3, APC, 4 rue Elsa Morante, F-75013 Paris, France}
\affil[b]{Sorbonne Universit\'{e}, Universit\'{e} Paris Cit\'{e}, CNRS/IN2P3, LPNHE, 4 place Jussieu, F-75005 Paris, France}
\affil[c]{Universit\'{e} Paris Cit\'{e}, CEA, CNRS/IN2P3, APC, 4 rue Elsa Morante, F-75013 Paris, France}
\affil[d]{Universit\'{e} Grenoble Alpes, CNRS/IN2P3, LPSC, 53 avenue des Martyrs, F-38026 Grenoble, France}
\affil[e]{Aix Marseille Universit\'{e}, CNRS/IN2P3, CPPM, 163 avenue de Luminy, F-13288 Marseille, France}
\affil[f]{Universit\'{e} Clermont Auvergne, CNRS/IN2P3, LPCA, 4 Avenue Blaise Pascal, F-63000 Clermont-Ferrand, France}
\affil[g]{SLAC National Accelerator Laboratory, 2575 Sand Hill Rd., Menlo Park, CA 94025, USA}
\affil[h]{National Astronomical Observatory of Japan, Chile Observatory, Los Abedules 3085, Vitacura, Santiago, Chile}

\authorinfo{Corresponding author e-mail: aboucaud@apc.in2p3.fr}

\begin{document}

\maketitle

\begin{abstract}
The Filter Exchange System of the LSSTCam at the Vera C. Rubin Observatory
is a critical subsystem enabling the Legacy Survey of Space and Time (LSST) by
performing rapid, repeatable exchanges among five large-format filters within a
highly constrained in-camera volume. Since the start of on-sky operations in
April 2025, the FES has routinely performed up to 40 filter exchanges per night,
completing each change in under \SI{90}{\second} with a positioning
repeatability of \SI{100}{\micro\meter} in the focal plane.
Safety and reliability are ensured through a dedicated software architecture.
Drawing on over a year of operational experience, we report on
the in-situ performance of this sophisticated system within the observatory environment,
characterize the key performance metrics, and discuss how specific design
choices have influenced system behavior and reliability in practice.
\end{abstract}



\section{Introduction}
\label{sec:intro}
The Filter Exchange System (FES; see Ref.~\citenum{Juramy22}) of the
LSST Camera (LSSTCam; see Ref.~\citenum{Lage24}) at the Vera C. Rubin
Observatory handles five \SI{75}{\centi\meter} diameter filters, each weighing between
\SI{25.5}{\kilogram} and \SI{38}{\kilogram}, for a total in-camera mass of
approximately \SI{160}{\kilogram}. 
The primary requirement of the FES is to perform a filter exchange in less than
\SI{90}{\second}, with a positioning accuracy of \SI{100}{\micro\meter} at the
focal plane, while operating within the very tight spatial constraints of the
camera body.

\begin{figure}
    \centering
    \includegraphics[width=0.85\linewidth]{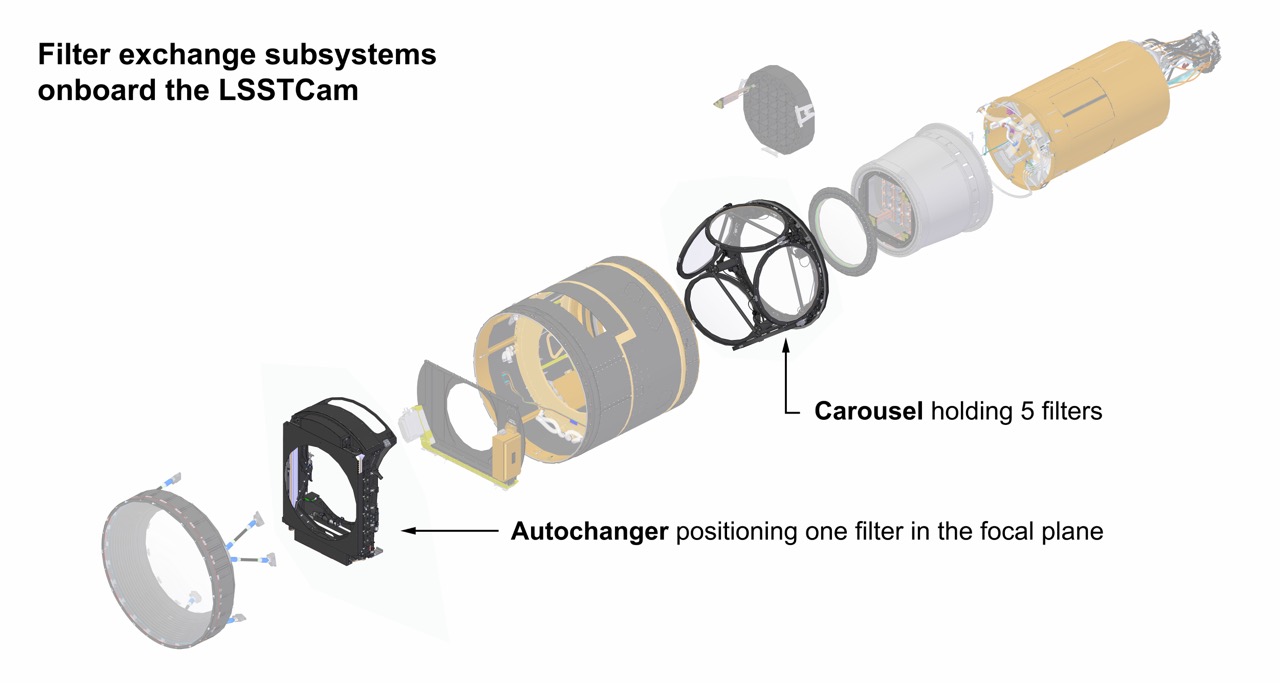}
    \caption{Exploded view of the LSSTCam with annotated elements of the filter exchange system: 
    the carousel and the autochanger. Credit for original image: T. Lange (SLAC)
    }
    \label{fig:fes-elements}
\end{figure}

\noindent The FES is composed of three motorized subsystems, each with a specific role:
\begin{itemize}
\item \textbf{The Carousel:} stores up to five filters outside the beam, in five
stands forming a pentagonal shape surrounding the cryostat, named sockets. The
carousel has one specific position called \textit{standby} that acts as the
interface with the autochanger. In order for a selected filter on the carousel
to be installed for operation, its socket must first be rotated to the
\textit{standby} position. Seen in Figure~\ref{fig:fes-elements}.
\item \textbf{The Autochanger:} transfers the filter from the carousel
\textit{standby} position to the focal plane along a complex 3D trajectory. It
then secures the filter in place with a \SI{100}{\micro\meter} accuracy. Seen in Figure~\ref{fig:fes-elements}.
\item \textbf{The Loader:} which can be mounted on the camera body during 
daytime operations, interacts with the autochanger to swap filters in and out
of the LSSTCam, enabling the use of the full six-filter LSSTCam set. Shown in Figure~\ref{fig:loader}.
\end{itemize}

At every operational stage, at least one of these subsystems manages the filter using a
dedicated locking mechanism. The design and qualification processes for the
FES addresses seismic risks, as well as the observatory’s
temperature range and the stringent cleanliness requirements for both filters
and lenses.
System safety is guaranteed by multiple Programmable Logic Controllers (PLCs).
The FES control is fully integrated into the Filter Control System
(FCS) software, which interfaces with all FES hardware, including 13 motors,
approximately 100 sensors, 22 controllers, 3 PLCs, and one PC104 unit—via a
robust CAN bus communication network.

After one year of on-sky operations, 3200+ successful filter changes
have been performed, with an average of 9 changes per night and up to 40 changes on some nights. This document presents the performance of the
LSSTCam Filter Exchange System (FES) and evaluates the impact of the design
choices made during its development.


\section{FES hardware during commissioning}
\label{sec:commissioning}

The FES is housed inside the LSSTCam camera body. This enclosed volume maintains
circulating air at \SI{0}{\percent} humidity to prevent condensation on the
large cryostat entrance window. The integrated LSSTCam camera body environment is drier than the
original FES design assumption, which had anticipated a humidity level of
approximately \SI{3}{\percent}.
Before installation on the telescope, the FES had never operated in such dry
conditions. Consequently, the primary hardware issues encountered during
commissioning were linked to the dry atmosphere that was previously unexplored during testing. Although these
issues have been resolved, it is worth documenting them and highlighting the inherent robustness of the FES design.

Two components of the FES required
adaptation during commissioning to address behavior driven by the
extremely dry environment of the LSSTCam camera body.

\subsection{Carousel clamp friction}
\label{sec:clamp-friction}
When stored in the carousel, each filter is supported at three points: one at
the bottom for guidance only, and two on the sides that secure the filter using
mechanical clamps. These clamps are self-locking, engaging automatically when
the autochanger pushes the filter’s side pins into them. Conversely, to allow
the autochanger to retrieve the filter, a motor-driven mechanism opens the
clamps whenever requested by the FCS. Motion is transmitted from the carousel’s fixed section to its rotating
section via a magnetic coupling and a cable.

Probes monitor both the clamp state (open/closed) and the filter’s position
within the clamp. Initially, it was expected that the moving parts inside the
clamps would operate reproducibly with minimal friction, even without grease.
The primary reason for avoiding grease was to reduce maintenance requirements,
as its application would necessitate removing the clamp from the camera and
disassembling it.

From the start of commissioning, operating for the first time under \SI{0}{\percent}
humidity, the clamps exhibited significant friction, as evidenced by a
noticeable increase in the time required to open them. After two months of
commissioning, it became necessary to revise the opening trajectory (motor current vs.
time) and increase the driving current. Simultaneously, an automatic
self-recovery procedure was put in place. This procedure (see
Section~\ref{sec:autorecovery}) involves briefly actuating the unclamping motor
to relieve tension in the unlocking cable, thereby facilitating proper clamp
closure.

This self-recovery procedure—automatically triggered by the slow control system
when clamp closure is not detected within \SI{2}{\second}, despite the filter
being in its expected position—was triggered with increasing frequency after its initial integration in September 2025. After
two months, the procedure had to be systematically applied to two of the ten
clamps. Ultimately, one clamp failed completely, reducing the number of usable
filters from five to four until maintenance could be performed. In October 2025, 
all clamps underwent maintenance:
\begin{itemize}
\item All clamps were disassembled and inspected.
\item The non-operational clamp exhibited significant friction during manual
operation. However, this friction decreased substantially after the clamp was
exposed for one day to a controlled environment at \SI{35}{\percent} humidity.
This suggests that humidity affects the surface properties of the moving parts.
\item Nevertheless, all clamps still exhibited some friction during manual
operation, even two days after removal from the dry environment of the camera. While
this low-level friction may have always been present, it is likely highly
dependent on surface properties and humidity.
\end{itemize}

\begin{figure}
    \centering
    \includegraphics[width=0.95\linewidth]{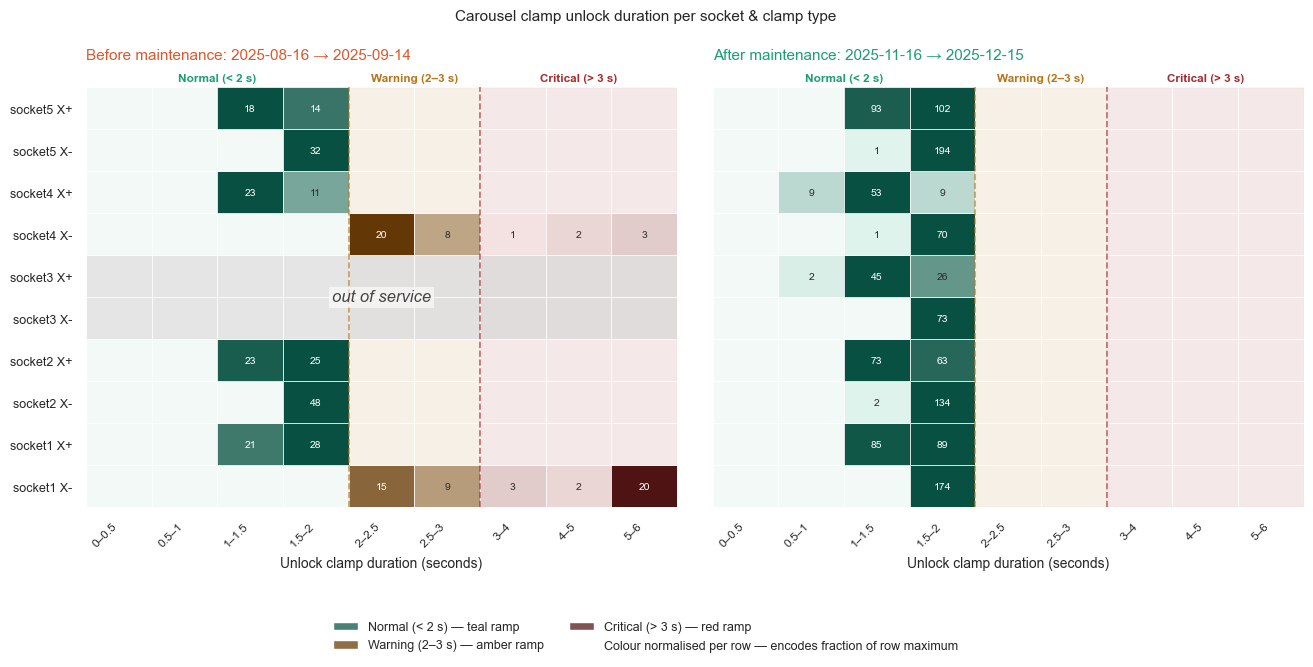}
    \caption{Set of heatmaps showing the \texttt{unlock} duration of all the carousel socket X+ and X- clamps before (left) and after (right) the October 2025 maintenance. We can see that the unlocking duration of the X- clamp of sockets 1 and 4 was particularly slow compared to the others, exhibiting signs of internal friction, that was fully solved after the lubrication during maintenance. Due to an unexpected hardware issue, socket 3 was deactivated mid-July 2025, which is why it does not appear on the left heatmap.}
    \label{fig:fes-unlock}
\end{figure}

To address this issue, a minimal amount of grease was applied to the main
sliding surfaces inside the clamps. The objective was to create a thin layer
ensuring consistent sliding properties, regardless of ambient humidity. After
the maintenance and a subsequent testing period, the current required to open the clamps was reduced by
approximately \SI{20}{\percent} (about \SI{1}{\ampere}). This intervention fully
resolved the issue: clamp operations are back to their nominal speed and are
highly reproducible (see Figure~\ref{fig:fes-unlock}).

%

These surface property changes caused by the extreme dryness inside the
camera enclosure had an additional unexpected effect on signal and current transmission to and from the rotating part of the
carousel.

\subsection{Carousel collector friction}
\label{sec:collectors}
The carousel rotates to position filters for the autochanger. Electrical power
and sensor signals from the rotating section are transmitted via five copper
slip rings on the back flange and carbon–silver brushes (called
\textit{collectors}) mounted on the rotating part. The brushes are spring-loaded
to ensure continuous contact with the copper rings.

\begin{figure}
    \centering
    \begin{subfigure}[b]{0.48\textwidth}
        \centering
        \includegraphics[width=0.9\textwidth]{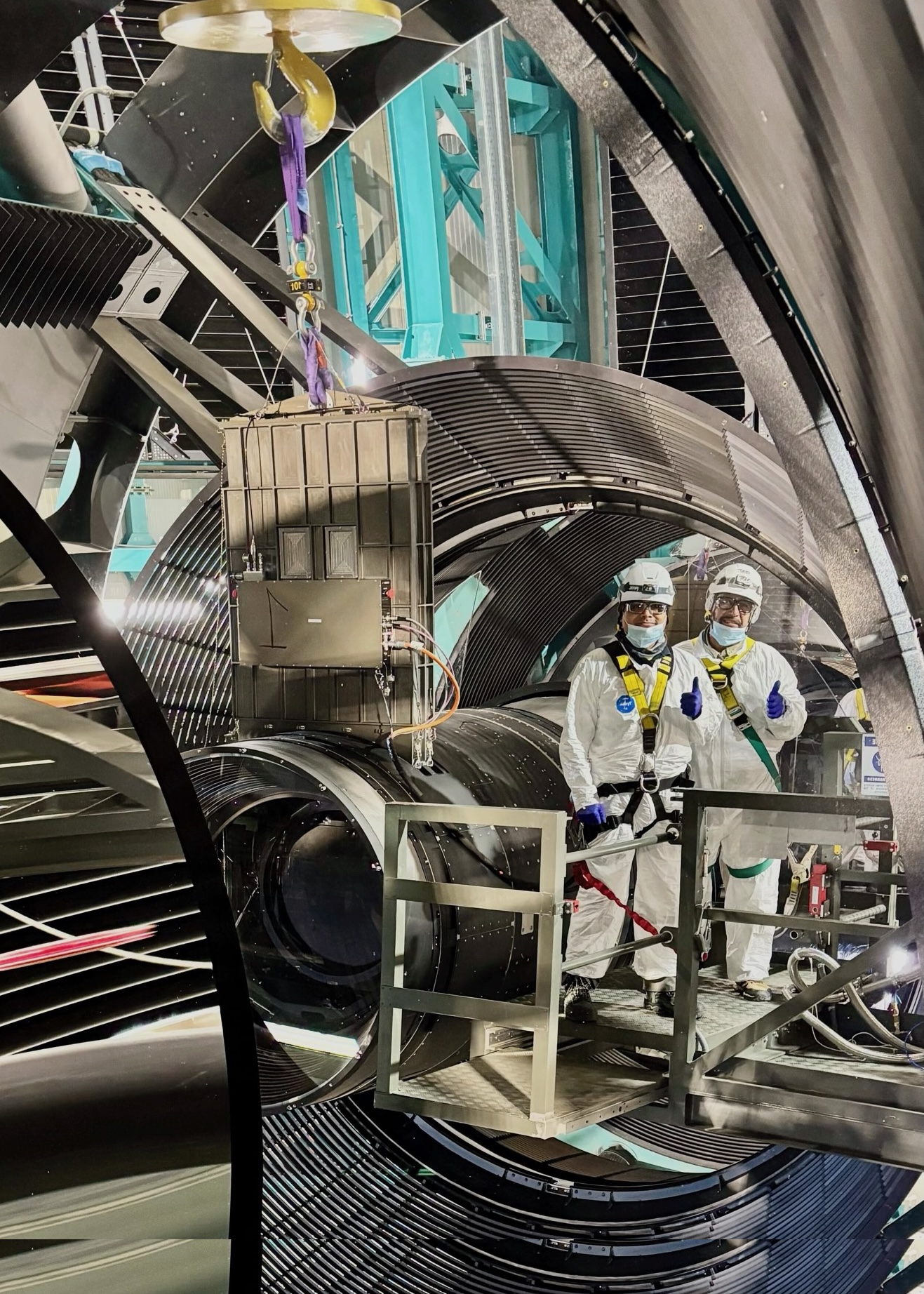}
        \caption{One of the two FES loaders, connected to the LSSTCam during a filter swap operation. Photo credit: T. Lange (SLAC)}
        \label{fig:loader}
    \end{subfigure}
    \hfill
    \begin{subfigure}[b]{0.48\textwidth}
        \centering
        \includegraphics[angle=90, width=0.8\textwidth]{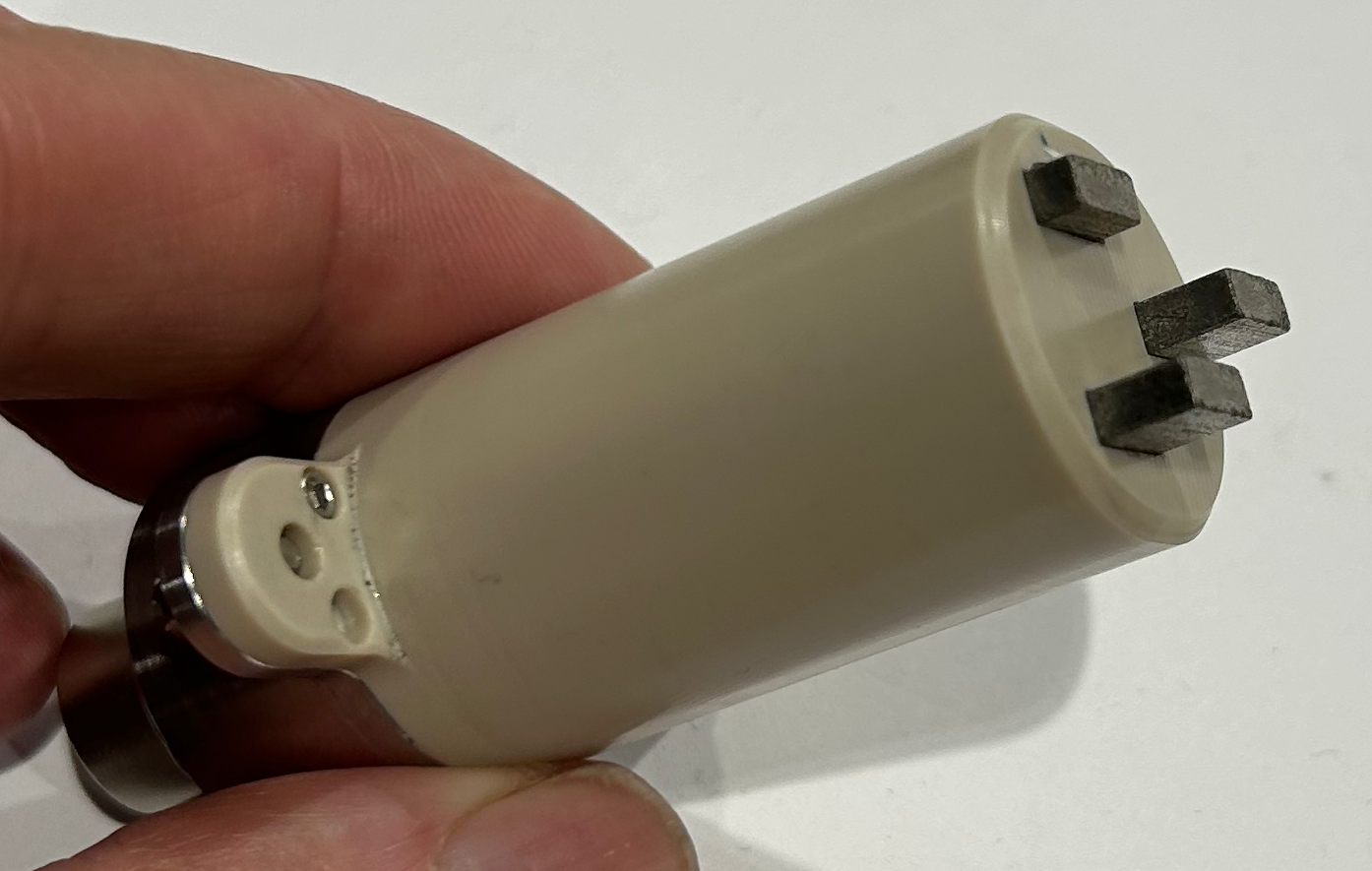}
        \caption{The faulty collector after removal from the carousel, with the left carbon–silver contact
        blocked in a ``spring compressed'' position due to friction. The other two carbon–silver contacts are fully extended in their nominal position.
        }
        \label{fig:collector}
    \end{subfigure}
    \vspace{0.3em}
    \caption{Photos showing the FES loader attached to LSSTCam, and the faulty carousel collector.}
    \label{fig:photos}
\end{figure}

At the start of operations in the dry atmosphere, communication errors related to
the slip-ring system started to appear. Each channel uses two collectors: one for power
transmission and the other for CAN bus signal transmission. In late April 2025,
however, one channel completely lost communication capability. Disassembly of
the faulty collector revealed that one carbon–silver pin was stuck and
non-operational (see Figure~\ref{fig:collector}). Notably, after approximately 24
hours in a $\approx$\SI{35}{\percent} humidity environment outside the camera, the residual
friction, present after manually freeing the contact, had disappeared.

In practice, the issue inside the camera gradually resolved itself over time:
system operation provided a beneficial burn-in, gradually smoothing the
increased surface roughness caused by the dry atmosphere. The error rate,
initially at zero, rose to over one communication glitch per rotation at the
start of operations in April 2025, and returned to zero by mid-May 2025.


\section{FES protection system}
\label{sec:plc}

\subsection{PLC and FES protection system}
Programmable Logic Controllers (PLCs) are designed to protect the filters
from incorrect or dangerous FES operations. Using input from probes and sensors
that continuously monitor the system state, the PLCs either enable or block
specific operations based on the current FES state. In real time, the PLCs
deactivate system components by controlling power to the motors via relays.

To prevent collisions between the filter and the carousel while the
autochanger is moving the filter, the logic follows:
\begin{itemize}
\item The relay powering the autochanger motors is enabled only when the
carousel is in a position and configuration compatible with autochanger
movement.
\item The carousel rotation controller does not receive the \SI{48}{\volt}
required to power its motor. Even if a rotation command is issued, it cannot be
executed.
\end{itemize}

This approach provides both a physical authorization for requested operations
and an automatic stop if the system enters an unexpected state. In addition to
the PLC, two further protection layers exist:
\begin{itemize}
\item An upper layer at the slow-control software level, which verifies that the requested operation is compatible with the current FES
state prior
to execution. Its primary purpose is to prevent human errors that would otherwise
trigger a PLC interruption, which would then require explicit error
acknowledgment and a controller reset before any new operation could proceed.
\item A lower layer at the motor controller level: these controllers are
configured to interrupt motion if the trajectory deviates excessively from
expectations, such as in the case of an unexpected obstacle or abnormal
friction.

\end{itemize}

The full set of PLC logic equations and behaviors was qualified prior to camera
installation on the telescope. However, during commissioning, we refined
some of the logic to minimize excessive relay activations during nightly telescope
operations.

\subsection{PLC upgrade during commissioning}
\label{sub:plc-upgrade}

\noindent The FES PLC commissioning included:
\begin{itemize}
\item Fine-tuning the allowed movement ranges based on position limit switch values, to
account for the final system configuration and the summit temperature range.
\item Adjustment of delays before PLC state changes to mitigate glitches
observed in some probes during FES operation.
\item Deactivation of the PLCs during non-FES operations.
\end{itemize}

The third adjustment was initially implemented only for the carousel PLC, as a
side effect of disabling carousel probes outside filter changes. This was
intended to reduce power dissipation within the camera. However, during commissioning, this PLC sleep mode had to be extended to the
autochanger PLC. Autochanger movements are permitted only within a narrow camera
rotator angle range (\ang{\pm5}), ensuring the autochanger rails remain aligned
with gravity, a requirement for proper filter displacement in the compact camera
environment. The carousel PLC monitors this orientation using an inclinometer
with a resolution of approximately \ang{0.4}.

During telescope tracking, the trajectory combined with inclinometer measurement
noise can cause high-frequency on/off switching of the autochanger motor relays.
To prevent premature relay aging, the PLC is deactivated when the FES is
idle, namely at the end of each filter change, and reactivated just before
the next operation begins.

This PLC sleep mode, particularly its ability to quickly restore full operation
at the start of a filter change, is detailed in Section~\ref{sec:wake-up}.

\subsection{FES protection system in action}

From April 2025 to May 2026, in addition to protecting the whole system against human error and routine operations that go awry, the
FES protection system executed several critical interventions, demonstrating its
capabilities.

Between April and December 2025, the PLC prevented FES operations 16
times, with no further occurrences afterward. Most of these vetoes occurred
during the core commissioning phase and were linked to non-optimal system
configurations or wake-up issues from sleep mode, which required time to
optimize (see Section~\ref{sec:wake-up}).

One notable event, related to the carousel clamp friction described in
Section~\ref{sec:clamp-friction}, is worth highlighting. Under normal
conditions (i.e., without friction), the clamp opens passively when a filter pin
enters it to reach its storage position. On this occasion, one carousel clamp blocked the
filter as its pins entered, preventing it from finishing the movement. 
This resulted in an asymmetric configuration: one
filter pin was partially inserted into its clamp, while the other remained stuck
outside. While the PLC temporarily allows this system configuration as a transition
state, it is only tolerated for a short period. After this delay, the system
triggered a shutdown of the autochanger motor power to prevent excessive and
potentially damaging force. Following this event, the PLC inhibited all
further FES operations.

Trending data—recorded at approximately \SI{10}{\hertz} during FES
operation and including the state of all probes and motors—made it possible to
diagnose the cause of the interruption (i.e., a clamp blocking the movement).
The camera was then positioned horizontally, and the PLC was manually overridden
to open the faulty clamp using a dedicated manual current ramp.
The PLC was reactivated, enabling safe storage of the filter in the carousel, and
maintenance of the faulty clamp was scheduled (see
Section~\ref{sec:clamp-friction}).

The PLCs, despite being programmable, are among the lowest-level layers of system programming.
Most updates are made to the slow-control layer called the Filter Control System (FCS).


\section{FES slow-control system: FCS}

The FES system control and operations rely on three software layers: 
\begin{itemize}
\item \textbf{Top level:} a Java-based system for communicating with the hardware
using the CAN bus protocol, executing filter exchanges, monitoring the entire
system, and hosting a graphical user interface for daily maintenance operations.
\item \textbf{Intermediate level:} the micro-code of the various motor
controllers combined with their configuration parameters that execute the
operations in the requested steps.
\item \textbf{Base level:} the protection system, PLC-based, that continuously ensures
filter safety.
\end{itemize}

The flexibility, complementarity, and robustness of this software design have
been demonstrated throughout commissioning. However, the high mechanical
complexity of the FES—which combines extreme spatial constraints with fast,
precise filter motion—makes the hardware inherently less flexible to modify than the software. While software optimizations are essential during
commissioning, key features—such as the mechanisms that securely hold the
filters without over-constraining them, or the 3D motion profile that limits
filter acceleration to $\lesssim$\,\SI{1.5}{\gee}—are complex and require careful
testing and qualification before deploying software changes to the camera that
could affect real filters.

\subsection{Software optimization during commissioning}

While the final qualification of slow-control software typically occurs during
on-site commissioning in Chile—where it competes with observatory operations—the availability of
a full-scale FES demonstrator and camera motion test bench at LPNHE laboratory in Paris enabled
independent integration, testing, and refinement prior to deployment on LSSTCam. This test infrastructure has been
essential for slow-control upgrades, optimization, and troubleshooting, enabling
detailed analysis of issues encountered during LSSTCam commissioning without
impacting observatory operations.

Another valuable resource during the commissioning period was the on-site presence of
Camera-FES team members at the observatory. In particular, on the software
side—which is often the last to receive development and qualification time—one
team member was full-time in Chile from October 2024 to November 2025 to bring
the FES control software (FCS) to operational readiness.

The rest of the section highlights key improvements made to the Filter Control
System (FCS) during commissioning that have significantly enhanced overall
system performance.

\subsection{Loader operation}

Changing the set of filters inside the camera requires the use of the filter loader
system. This device (seen in Figure~\ref{fig:loader}), which inserts and
extracts filters from the camera, must operate safely under the control of the
FES PLC.

In practice, all loader electrical and software connections have been
implemented as a ``hot-plug'' operation. The loader, only used during daytime
operations for loading and unloading filters, can be dynamically connected to
and disconnected from the camera, the power supply, the protection system and the FCS.

Because the power line cannot, by design, be de-energized at the camera
level, dedicated protections have been implemented on the loader side to
eliminate any risk during cable connection. The PLC logic accounts for the
presence or absence of the loader and extends its functionality accordingly when
the loader is connected.

A dedicated CAN bus is used for the loader, fully independent of the
carousel/autochanger CAN bus. It operates over longer cables and therefore at a
lower bitrate than the rest of the FES system (\SI{125}{\kilo\bps} versus
\SI{1}{\mega\bps}).

The software library used to manage the CAN bus on the FES PC104 successfully
supports hot-plug connection, initialization, and disconnection of the loader,
without requiring a restart of the FCS software. The entire operation can be 
handled by a software operator from a graphical user interface which displays 
live the status of the various sensors, the power supply switches, and two buttons:
one to extract a filter from the camera into the loader, and one to insert a filter from the loader into the camera.

The sequencing of the underlying hardware operations has been carefully tuned
and tested over a few dozen live operations to minimize blocking points, and
this full hot-plug capability has now been in operation for more than six months
without a single issue.

\subsection{FES integration within observatory control}

\subsubsection{FES degraded mode}
\label{sec:degraded}

To enhance system robustness and safety during interaction with the
observatory, we implemented a degraded mode that allows the camera to continue
operating seamlessly after acknowledging that some components of the FES are unavailable. It
prevents the system from entering a fault state when the FES
functionality is partially compromised. This ensures uninterrupted imaging under
suboptimal conditions, whether due to daytime checkout results or in-night
failures. The mode accommodates several scenarios, including:
\begin{itemize}
    \item Unusable carousel sockets, rendering specific filters unavailable.
    \item Inoperable carousel or autochanger, restricting use to the currently
    installed filter.
    \item Rotation failure, limiting operation to the currently loaded filter, including no filter.
\end{itemize}

Degraded mode can be activated via CCS engineering mode, either at the start
of the night or dynamically in response to failures. Depending on the above
degraded scenario, it will immediately forward the list of currently available filters to the
observatory, meaning the list of installed filters in the camera
excluding those that cannot safely be placed in the focal plane.

\subsubsection{FES sleep mode / wake up}
\label{sec:wake-up}

To optimize energy efficiency within the LSSTCam, the FES carousel supports a
low power mode (also referred to as sleep mode), reducing consumption during
idle periods and avoiding heat buildup inside the camera body. While this mode
introduces a $\approx$\,\SI{5}{\second} overhead to wake up the system before
executing a filter change, this cost is small compared to the energy savings it provides. Additionally,
this overhead can fully disappear if the wake up is executed in anticipation
of a filter change, though this has not yet been implemented at the observatory level.

Development of this feature spanned several months, with incremental
improvements to both backend and user interface. In early 2025, the work focused on
core power state logic and integration with primary commands. By mid-2025, we
introduced automated power-saving after filter changes as well as checks to
prevent a ``bad wake up'', meaning FES actions executed while the system is
not fully responsive, leading to system alerts. Finally, we made this
functionality available at the observatory level so it can be controlled by
operators outside of the camera software. 

\noindent We thus offer three operational modes:
\begin{itemize}
\item \textbf{SLEEP}: enforces low power mode.
\item \textbf{WAKEUP}: transitions to high power mode in anticipation of a
filter change, automatically reverting to low power afterward.
\item \textbf{STAYUP}: maintains carousel power indefinitely, ideal for maintenance
or high-frequency operations.
\end{itemize}

\subsubsection{Live filter monitoring}
\label{sec:filter-monit}

The communication between the observatory and the camera subsystems is layered,
with the observatory control system written in Python and the entire camera
control system written in Java. The communication layer built on top
ensures fast information flow and keeps the observatory constantly aware of the
status of every camera subsystem.

Because the FES is one such subsystem, an important task was to ensure that
the full system status—particularly the filter currently in place—is accessible
to the observatory, guaranteeing that the recorded metadata remains valid at all
times during operations. To this end, all camera subsystems define two main
states: \texttt{NORMAL} and \texttt{ENGINEERING}. The \texttt{ENGINEERING} mode
indicates that the camera team is manually working on the subsystem and is in
charge, signaling to the observatory that the camera is unavailable for
operations. To enter operational mode, every camera subsystem must first switch
to \texttt{NORMAL} mode.

The transition from \texttt{ENGINEERING} to \texttt{NORMAL} state for the FES
evolved as the team gained confidence in the system’s reliability, reaching a
level where we require a filter to be placed in the focal plane and ready for
observations before this transition is completed. In practice, this means the
system is returned to the observatory in a state where sky images can be
captured immediately, without requiring a double-check of the subsystem’s
status.

\subsection{Auto-recoveries}
\label{sec:autorecovery} 

The implementation of auto-recovery procedures has both improved the reliability
of FES operations—ensuring uninterrupted execution—and significantly enhanced
repeatability. In particular, FES motion execution now demonstrates reduced
timing jitter.

Given the high speed and complexity of FES operations, some mechanical
variability is expected over time. The system experiences a wide range of forces
and constraints, depending on filter mass, telescope orientation, operational
conditions, and interactions between subsystems (controllers, motors, and
mechanisms).

The guiding principle of these auto-recovery procedures is to leverage the
extensive sensor network and real-time system state monitoring to detect
out-of-tolerance behavior and take immediate corrective action, restoring the
system to its optimal operating state. These procedures can also recover from
failures, including controller errors, by automatically resolving the underlying
issue and resuming normal operation without human intervention.

\noindent Below are two examples of auto-recovery procedures implemented in the FCS:
\begin{itemize}
\item \textbf{Carousel clamp unlocking:} friction in the clamp mechanism can
prevent timely opening when using the nominal current. However, a reconfigured
actuation mode that applies higher current can overcome this condition. Although
this ``slow mode'' increases operation time and accelerates system wear, it is
used selectively as an auto-recovery mechanism. This allows operations to
continue while maintenance (e.g., re-greasing) is scheduled at a convenient
time,
\item \textbf{Filter load force relaxation:} when a filter mounted in the loader
is transported within the dome using the crane, mechanical constraints can build
up in the hook mechanism that holds the filter. As the filter begins to move
down into the camera, the PLC may interpret these forces as unexpected motion,
triggering an interruption. The auto-recovery procedure detects this pre-load
condition before motion begins, releases the constraint, and then safely
initiates the in-camera filter loading sequence.
\end{itemize}


\section{Performance}
\subsection{Number of filter changes}

During the LSST survey operations, we expect a rate of filter changes up to
$\approx$17 per night. To this rate, we should add 1 to 5 filter changes during
daytime to take into account telescope calibration and FES daily operations
for maintenance or filter swap. 

During this year of commissioning, from the 14th of April 2025 to the 14th of
April 2026, we counted:
\begin{itemize}
    \item a total of 3222 filter changes (averaging to $\approx$9 filter changes a day),
    \item 85 days without a filter change (maintenance, AOS tuning, bad weather),
    \item 197 days with 5 filter changes or more,
    \item 132 days with 10 filter changes or more,
    \item 49 days with 20 filter changes or more,
    \item 71 filter changes – maximum recorded during a single night, associated
    with the commissioning of the filter focus offset.
\end{itemize}

To avoid overheating of the FES autochanger motors, a \SI{2}{\minute} pause
between two filter exchanges is requested, limiting the maximum rate to 15
filter changes per hour during normal operations. For specific targets of
opportunity, this limitation can be ignored for a maximum of $\approx$5 successive
filter changes in order to keep the increase of the FES autochanger motor
temperature below \SI{20}{\degreeCelsius}.


\subsection{Filter change duration and associated telescope overhead}
\label{sec:duration}

The time used to perform a filter change is an overhead to the light
collection, and has to be minimized. The design goal for the execution time of a
filter change is \SI{90}{\second} for the filter change itself by the FES to
which we should add \SI{30}{\second} to set up the telescope in the right
configuration and to go back to normal operation. 

\begin{figure}
    \centering
    \includegraphics[width=0.95\linewidth]{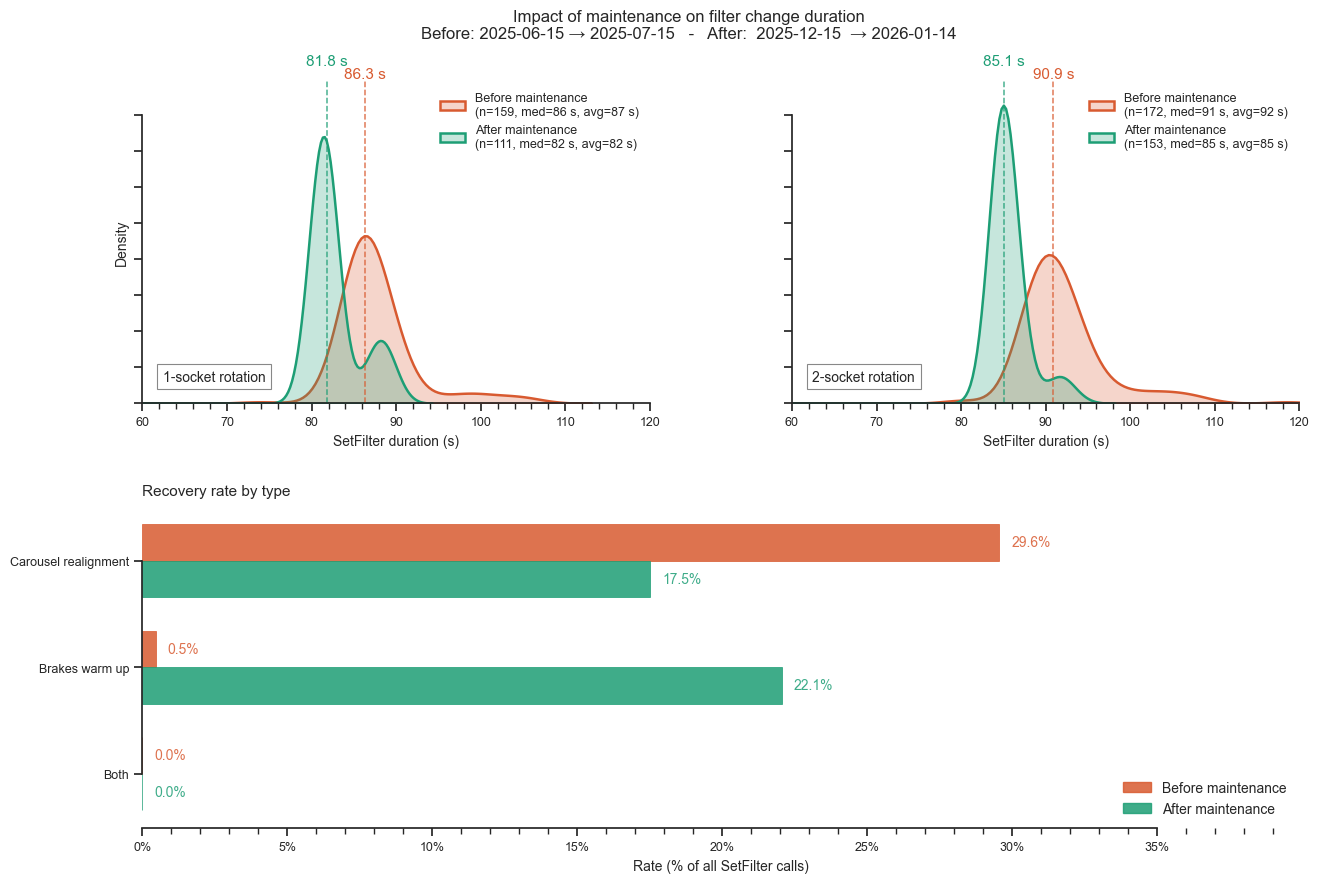}
    \caption{Top panels: Kernel density estimates of the filter change (\texttt{SetFilter} calls) duration distribution split by carousel rotation distance (1-socket and 2-socket) for each period. Median durations decreased after maintenance in both cases (1-socket: \SI{86.3}{\second} to \SI{81.8}{\second}; 2-socket: \SI{90.9}{\second} to \SI{85.1}{\second}), and the distributions narrowed, suggesting more consistent filter-exchange performance overall. Bottom panel: A bar chart of automated recovery rates (as a percentage of all \texttt{SetFilter} calls) explaining the tails or second peaks in the above distributions. We list here carousel realignment, carousel brakes warm up, and both combined. The carousel realignment rate decreased from \SI{29.6}{\percent} to \SI{17.5}{\percent}, indicating improved carousel balance post-maintenance. However, the carousel brakes warm up rate rose from \SI{0.5}{\percent} to \SI{22}{\percent}, reflecting an issue that emerged from the low number of filter exchanges during the period examined and the recent adoption of this recovery method. Fortunately, these recoveries never seem to happen during the same operation. Overall, the filter exchange duration is now well below the system requirement of \SI{90}{\second}.
    }
    \label{fig:fes-setfilter_jitter}
\end{figure}

As shown in Figure~\ref{fig:fes-setfilter_jitter}, the mean execution time of the
FES filter change is between \SI{82}{\second} and \SI{85}{\second} depending on the distance the carousel has to travel to fetch the next filter. In both cases we are well below the requirement after the
carousel clamp friction underwent maintenance (see Section~\ref{sec:clamp-friction}).

The main dispersion in the filter exchange time corresponds to the pentagonal
geometry of the FES carousel leading to two given filters being separated by
either one or two slots in any direction. Hence, during a filter change, the
carousel must either be rotated by 1/5 or 2/5 turn, leading to the two plots
at the top of Figure~\ref{fig:fes-setfilter_jitter}.

On top of these peaks, a small tail visible in
Figure~\ref{fig:fes-setfilter_jitter} is associated with two other effects:
\begin{itemize}
    \item \textbf{Carousel brakes warmup (\SI{5}{\second}):} if the carousel has
    not been used for more than \SI{5}{\hour}, its brakes must be operated
    independently to ensure rapid opening during normal use. This addresses a
    sticky contact issue that occurs when the brakes remain closed for extended
    periods. This process takes $\approx$\SI{5}{\second} at the start of each night
    to warm them up.
    \item \textbf{Carousel realignment (\SI{3}{\second}):} the carousel brakes
    are in contact with the main gear tooth, but due to backlash between the brake
    and the gear tooth, the gear can still move by approximately
    \SI{200}{\micro\meter}, especially if the telescope moves or the camera
    rotates. The tolerance we impose for the carousel-autochanger offset is
    below~\SI{100}{\micro\meter}. If this misalignment is detected prior to a filter
    change, the carousel automatically performs a small realignment with the
    autochanger before storing the filter, which takes \SI{3}{\second}.
\end{itemize}

The overhead associated with moving the camera rotator to \ang{0} and back to
its target value is \SI{0.27}{\second\per\degree}, in addition to a constant
overhead of \SI{32}{\second}. For an average \ang{\pm45} rotation, this results
in a dead time of \SI{57}{\second} for the telescope operation associated with a
filter exchange. At this stage of commissioning, the total dead time associated
with a filter change is \SI{141}{\second}, that is \SI{21}{\second} longer than
the target. There is still room for optimization that has not yet been
implemented, due to other priorities in observatory optimization and
commissioning.
\begin{itemize} 
    \item Currently, telescope tracking is stopped
    and restarted during a filter change. This protection configuration, set during commissioning,
    should be removed over time after testing on-summit.
    \item While the camera rotator must be at zero for a filter change, LSST survey strategy can minimize the required rotation at the time of
    the filter change by requesting a filter change when the rotator is closest to \ang{0}.
    \item The duration quoted above for the pure-FES filter change does not include the FES
    wake-up time ($\leq \SI{5}{\second}$, see Section~\ref{sec:wake-up}). This
    operation can be performed in parallel with the telescope setup, prior to
    the start of the filter exchange at the FES level. This is not yet activated
    at the telescope control level; however, this duration is included in the
    telescope overhead, not the FES overhead.
\end{itemize}


\subsection{Filter positioning repeatability}

Given the constrained available space and the large size of the LSSTCam filters,
the designed filter exchange solution involves large movements, complex 3D
trajectories, and relatively high speeds. For example, the autochanger’s main
movement to transport the $\approx$\SI{30}{\kilogram} filter from the focal
plane to its carousel socket takes \SI{6}{\second} to cover
\SI{91}{\centi\meter} (excluding the \SI{9}{\centi\meter} of slow-speed approach
movement), with a \ang{90} change in direction. Despite this complexity, the stability of the instrument’s
photometric response is paramount. Although the spatial uniformity of LSST
filters is very good, it is not perfect, and even in a controlled, enclosed
environment, the presence of dust particles on the filters cannot be entirely
excluded. Therefore, the filter position in the plane transverse to the optical
axis must be reproducible with an acceptable tolerance of
\SI{100}{\micro\meter}, in order to preserve LSSTCam optical performance.

In the X–Y plane, this reproducibility is achieved through the FES autochanger
clamping system. Positioning in the X direction is
directly driven and monitored by the autochanger’s dual-arm trucks and their
encoders. This achieves a reproducibility of approximately
\SI{50}{\micro\meter}—an order of magnitude better than required. In the Y
direction, dedicated position sensors monitor and ensure consistent filter
presentation to the focal plane clamps, regardless of telescope pointing and
gravity effects.

\begin{figure}
    \centering
    \includegraphics[width=0.90\linewidth]{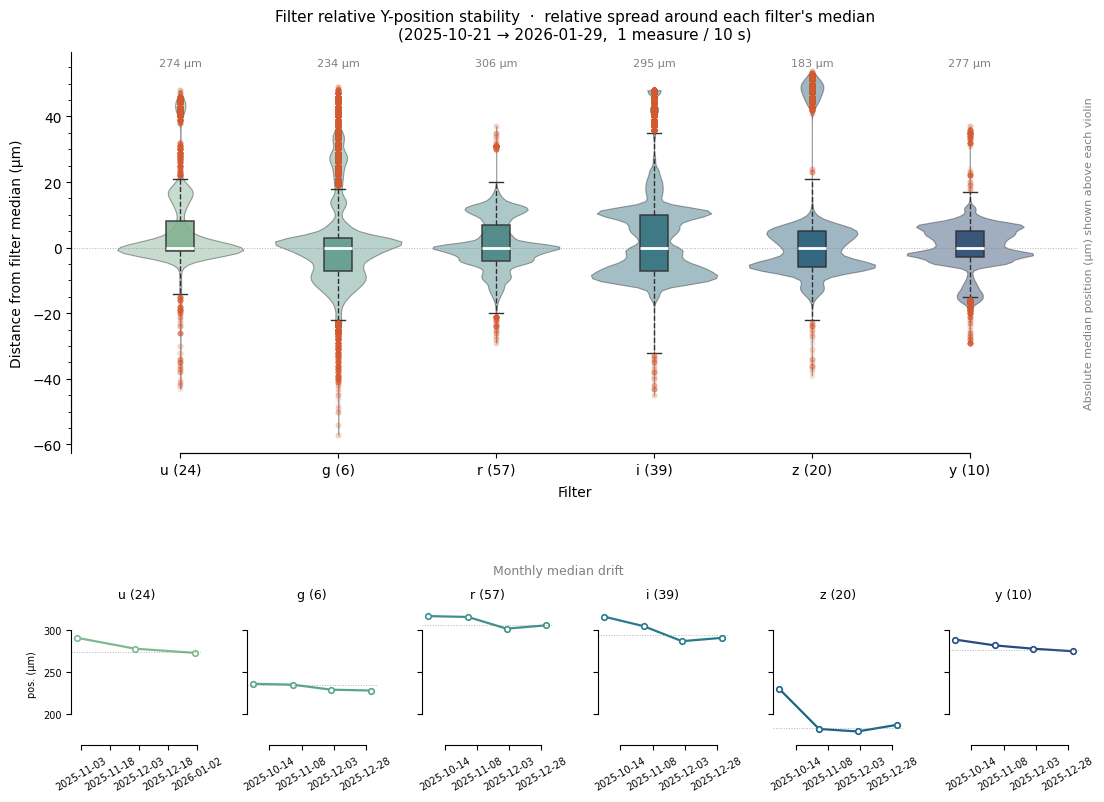}
    \caption{The
    positional reproducibility over the three-month period is shown, with peak-to-peak
    within $\pm$\,\SI{100}{\micro\meter}, across all filters. In practice,
    the observed dispersion is associated with position changes at each filter
    exchange, as the monitoring indicates displacement below
    \SI{1}{\micro\meter} once the filter is in place.
    }
    \label{fig:yposition}
\end{figure}

While the absolute positioning may vary slightly from one filter to another due
to small differences in their mechanical interfaces with the clamping system,
the measured positioning reproducibility remains below \SI{100}{\micro\meter} peak-to-peak, as
illustrated for the Y direction in Figure~\ref{fig:yposition},
a factor of three better than required.


\subsection{Swapping filters in/out of LSSTCam}

Among the six LSSTCam filters, \{\textit{u, g, r, i, z, y}\}, the camera body
can only hold five simultaneously, resulting in a need for what we call filter
swaps. During a filter swap, one of the two FES loaders extracts one of the five filters
out of the camera, and the other FES loader inserts the sixth filter. While we could imagine a turnover
between all six filters, in practice, only the filters at both ends of the
spectrum, \textit{u} and \textit{y}, are swapped alternately every other week,
corresponding to moon cycles. Over the past year, in addition to routine
\textit{u}/\textit{y} swaps, the loaders have also been used to install
technical filters such as the pinhole filter or the empty frame, and to insert
or remove filters for safe maintenance access.

The filter swap procedure takes about two hours and involves the two FES loaders
(each weighing $\approx$\SI{200}{\kilogram} with a filter inside). One is used to
extract a filter from the camera, and the second to install a new one. Most of
this operation is spent moving securely and sequentially the loaders from their
chariot on the dome floor to the camera on the telescope, using the dome crane
(\SI{20}{\minute} each way). Once removed from the camera, the unused filter can
stay inside a loader, ready for the next swap operation. Due to ongoing
construction activities in the dome, the loaders are currently stored in the
observatory clean room. In the future, they will remain inside the dome
and be connected to a clean airflow system.

A filter swap operation requires four persons in radio contact the whole time: a
crane operator, two mechanical operators whose role is to secure the interface
between the camera and the FES loaders (opening and closing panels, cleaning the
dust, managing the external power cables and CAN bus lines), and one software
operator that controls the FES via the Filter Control System (FCS). The
qualification of the swap procedure, the full automation of FES operations, and
team training were successfully completed by November 2025. 

Over one year (April 2025 to April 2026), we report:

\begin{itemize}
\item 54 filter load or unload operations (including operations to/from a
filter storage box in the clean room),
\item 11 \textit{u}/\textit{y} filter swaps (23 swaps per year are expected during
normal operations),
\item 16 additional filter swaps requested for maintenance ($\leq$5 per year
expected during normal operations).
\end{itemize}


\subsection{Error rate evolution from April 2025 to April 2026}

In the first six months of the commissioning, the error rate requiring human
intervention decreased from \SI{5}{\percent} to $\lesssim$\,\SI{0.1}{\percent} of
filter changes. Issues encountered have been mitigated by 
\begin{itemize} 
\item FES, probes and PLC optimization, driven by operating the FES
in the observatory environment.
\item Software, mainly by writing/extending auto-recovery methods (see Section
\ref{sec:autorecovery}).
\item Hardware interventions on the camera in May and October 2025, to address
early-life issues in the system (see Sections \ref{sec:clamp-friction} and
\ref{sec:collectors}).
\end{itemize}


\section{Conclusion}

After one year of on-sky operations from April 2025 to April 2026, the LSSTCam
Filter Exchange System has demonstrated reliable, high-performance operation
within the Vera C. Rubin Observatory. Over this period, 3222 filter changes were
executed, up to 40 per night, with a mean execution time of \SI{84}{\second}, well
below the \SI{90}{\second} requirement.

Commissioning revealed unexpected hardware challenges, primarily driven by the
\SI{0}{\percent} humidity environment inside the camera enclosure. These affected
the carousel clamp and slip-ring collector mechanisms. Both issues were
successfully diagnosed and resolved through targeted hardware interventions—in
particular, re-greasing the clamp sliding surfaces—and complemented by software
auto-recovery procedures that maintained operations in the interim with minimal
human intervention.

On the software side, the Filter Control System underwent significant
development, delivering key capabilities including FES degraded mode for graceful
degradation, a PLC sleep/wake cycle to reduce power dissipation, hot-plug loader
integration, and automated self-recovery procedures. These improvements drove the
FES error rate requiring human intervention from \SI{5}{\percent} at the start of
operations to below \SI{0.1}{\percent} within six months.

The commissioning of the FES was formally declared complete in December 2025.
Ongoing efforts focus on long-term monitoring and stability, in particular
leveraging the full-scale FES demonstrator in Paris to study and qualify future
upgrades before deployment—an asset that has proven invaluable throughout
commissioning. The system is now well positioned to support the full duration of
the LSST survey.


\acknowledgements

This material is based upon work supported in part by the National Science Foundation through Cooper-
ative Agreements AST-1258333 and AST-2241526 and Cooperative Support Agreements AST-1202910 and
AST2211468 managed by the Association of Universities for Research in Astronomy (AURA), and the De-
partment of Energy under Contract No. DE-AC02-76SF00515 with the SLAC National Accelerator Laboratory
managed by Stanford University. Additional Rubin Observatory funding comes from private donations, grants
to universities, and in-kind support from LSST-DA Institutional Members. 
This work has been supported by the French National Institute of Nuclear and Particle Physics (IN2P3) through dedicated funding provided by
the National Center for Scientific Research (CNRS). The authors thank Sean MacBride for the thoughtful review of the manuscript.

\bibliography{report} 
\bibliographystyle{spiebib} 

\end{document}